\documentclass[superscriptaddress,nofootinbib,notitlepage]{revtex4-1}
\usepackage{amsfonts}
\usepackage[bookmarksopen,colorlinks]{hyperref}
\usepackage{graphicx}
\usepackage{dcolumn}
\usepackage{bm}
\usepackage{amssymb,amsmath}
\usepackage[dvips]{color}

\def\bf{\textbf}

\begin{document}

\title{Reflections on the covariance of modified teleparallel theories of gravity}
\author{Cecilia Bejarano} \email{cbejarano@iafe.uba.ar} \affiliation{Instituto de Astronom\'ia y F\'isica del Espacio (IAFE, CONICET - Universidad de Buenos Aires),  Casilla de Correo 67, Sucursal 28, 1428~Buenos~Aires,~Argentina} 
\author{Rafael Ferraro}  \email{ferraro@iafe.uba.ar} \affiliation{Instituto de Astronom\'ia y F\'isica del Espacio (IAFE, CONICET - Universidad de Buenos Aires),  Casilla de Correo 67, Sucursal 28, 1428~Buenos~Aires,~Argentina} \affiliation{Departamento de F\'isica, Facultad de Ciencias Exactas y Naturales, Universidad de Buenos Aires, Ciudad~Universitaria, Pabell\'on I, 1428~Buenos~Aires, Argentina}
\author{Franco Fiorini} \email{francof@cab.cnea.gov.ar} \affiliation{Departamento  de  Ingenier\'ia  en  Telecomunicaciones  and  Instituto  Balseiro, Centro  At\'omico  Bariloche,  Av.~Ezequiel  Bustillo  9500, CP8400  San Carlos de  Bariloche, Argentina} 
\author{Mar\'ia Jos\'e Guzm\'an} \email{maria.j.guzman.m@gmail.com} \affiliation{Departamento de F\'isica y Astronom\'ia, Facultad de Ciencias, Universidad de La Serena, Av. Juan Cisternas 1200, 1720236 La Serena, Chile}

\begin{abstract}
{\ We review the
current status of the Lorentz covariance in teleparallel and
modified teleparallel theories of gravity, and discuss the
controversial features of the different approaches. We also revisit
the issue of the remnant Lorentz gauge symmetries in $f(T)$
gravity.}
\end{abstract}


\maketitle

\section{Introduction}

The dynamical equations of any physical theory can always be found to be
invariant with respect to a~group of transformations, commonly termed as a
covariance group. The principle of covariance in general relativity
refers to the invariance of the form of the equations of motion under
differentiable coordinate transformations, also known as the diffeomorphism
group \cite{Sun14}. In the context of gravity theories written in terms of
frame fields (also known as vierbein or tetrads, in dimension four), there
is an additional symmetry group related {to} the freedom {of boosting or rotating}
such frames. Since tetrads can be considered as locally mapping spacetime
points in the tangent space and vice versa, they possess additional free
indices related to the tangent space. Global and local Lorentz
transformations in the tangent space indices are then allowed; therefore it can be talked about an additional covariance under Lorentz transformations in
these kinds of theories, and discuss whether the action and the dynamical
equations are covariant. It is clear that this discussion does not appear in
theories written in terms of the metric tensor, as they are formulated in
Lorentz covariant form from first principles.

Theories of gravity written in terms of the tetrad field have been
considered since the beginning of general relativity. Early attempts to
include the tetrad and the spin connection as independent variables in a
gravitational theory were considered in Einstein--Cartan gravity; these are
gravitational theories in a~Riemann--Cartan manifold with non-trivial
curvature and torsion. On the other hand, it is possible to achieve a~fully
equivalent version of general relativity in the tetrad formalism through the
consideration of a~curvatureless spin connection. These theories are built
in a~Weitzenb\"{o}ck spacetime, and~historically the simplest choice for the
spin connection is the Weitzenb\"{o}ck one, which depends linearly on
first-order derivatives of the tetrad field. This version of general
relativity is known as the teleparallel equivalent of general
relativity (TEGR), since the Weitzenb\"{o}ck connection has the property
that it parallel transports the tetrad field along the manifold, giving an
absolute notion of parallelism at a~distance \cite{Ald13}. The~Lagrangian of
TEGR is invariant under global Lorentz transformations of the tetrad field;
however, it is just pseudo-invariant if the Lorentz transformations become
local, since it differs from the Einstein--Hilbert Lagrangian
through a~boundary term that is sensitive to local Lorentz transformations.
This feature is not relevant for the equations of motion, as the boundary
term is integrated out once in the action, giving a~dynamics fully
equivalent to general relativity (GR). In particular, in spite of the fact that the tetrad contains more
components than the metric, TEGR only involves those degree{s} of freedom
associated {with} the metric \cite{Ferraro2016}.

Modified teleparallel theories of gravity where the TEGR is the underlying
framework are extensively being studied nowadays, with special emphasis in
confronting the main problems in cosmology related to the early
inflationary stage and the late accelerated expansion regimes. Nonlinear
modifications of a~Lorentzian pseudo-invariant Lagrangian are expected to
break local Lorentz symmetries, and to harbor extra degrees of freedom \cite%
{Ferraro:2018tpu}. Recently, several arguments have been put forward in favor
of covariant versions of modified teleparallel gravity that drop the
assumption of the Weitzenb\"{o}ck connection {in favor of} a~more general inertial spin connection, with its own advantages and problems. The purpose
of this article is to discuss the issues emerging from the peculiar behavior
of teleparallel theories under local Lorentz transformations of the tetrad.

\section{Teleparallel gravity}

\subsection{Geometric foundations of general relativity}

There are two main geometric objects that serve to build Lagrangians for
gravitational theories: the~torsion $\mathbf{T}^{a}$ and the curvature $%
\mathbf{R}_{\ b}^{a}$. The torsion is a~set of 2-forms defined as
\begin{equation}
\mathbf{T}^{a}=D\mathbf{E}^{a}=d\mathbf{E}^{a}+\boldsymbol{\omega }_{~\
b}^{a}\wedge \mathbf{E}^{b}.  \label{2ftorsion}
\end{equation}%

Here, the 1-forms $\mathbf{E}^{a}(x)$ make up the local basis of the
cotangent space ($\mathbf{E}^{a}=E_{\mu }^{a}~dx^{\mu }$), and $d$ is the
exterior derivative of forms. The set of 1-forms $\boldsymbol{\omega }_{~\
b}^{a} $ is the \textit{spin connection} that provides the exterior
derivative $D$ with a~covariant character under local Lorentz
transformations $\mathbf{E}^{a^{\prime }}=\Lambda _{~~a}^{a^{\prime }}(x)~%
\mathbf{E}^{a}$. This means that the tangent-space index in $\mathbf{T}^{a}$
is a~Lorentz vector index, in the sense that $\mathbf{T}^{a}$ transforms
like $\mathbf{E}^{a}$. For~this, the~spin connection $\boldsymbol{\omega }%
_{~\ b}^{a}$ must transform as%
\begin{equation}
\boldsymbol{\omega }_{\ b^{\prime }}^{a^{\prime }}=\Lambda _{\ a}^{a^{\prime
}}~\boldsymbol{\omega }_{\ b}^{a}~\Lambda _{\ b^{\prime }}^{b}+\Lambda _{\
a}^{a^{\prime }}~d\Lambda _{\ b^{\prime }}^{a},\label{omega}
\end{equation}%
where $\Lambda _{\ a^{\prime }}^{a}$ is the respective inverse Lorentz
transformation:\ $\Lambda _{\ a}^{a^{\prime }}\Lambda _{\ b^{\prime
}}^{a}=\delta _{\ b^{\prime }}^{a^{\prime }}$.

On the other hand, the curvature $\mathbf{R}_{\ b}^{a}$ is a~set of 2-forms
that depends only on the spin connection, and~reads
\begin{equation}
\mathbf{R}_{\ b}^{a}=d\boldsymbol{\omega }_{\ b}^{a}+\boldsymbol{\omega }_{\
c}^{a}\wedge \boldsymbol{\omega }_{\ b}^{c}.  \label{gralcurvature}
\end{equation}%

Like the torsion, the curvature behaves as a~Lorentz tensor under local
Lorentz transformations of the~basis.

The basis $\{\mathbf{e}_{a}\}$ of the tangent space ($\mathbf{e}%
_{a}=e_{a}^{\mu }~\partial _{\mu }$) is said \textit{dual} to $\{\mathbf{E}%
^{a}\}$\ if it fulfills $e_{a}^{\mu }~E_{\mu }^{b}=\delta _{a}^{b}~$, $%
e_{a}^{\mu }~E_{\nu }^{a}=\delta _{\nu }^{\mu }~$. In a~space endowed with a
metric tensor $\mathbf{g}$, it is useful to restrict the bases to be
orthonormal:%
\begin{equation}
\mathbf{e}_{a}\cdot \mathbf{e}_{b}=g_{\mu \nu }~e_{a}^{\mu }~e_{b}^{\nu
}=\eta _{ab},  \label{ortho}
\end{equation}%
where $\eta _{ab}=\mathrm{diag}(1,-1,-1,-1)$ is the Minkowski tensor. The
orthonormality condition allows for obtaining the metric from the tetrad; in
fact, by using the duality, one obtains%
\begin{equation}
g_{\mu \nu }=\eta _{ab}~E_{\mu }^{a}~E_{\nu }^{b}.
\end{equation}

In addition, the non-metricity tensor $Q_{\lambda \mu \nu }=\nabla
_{\lambda }g_{\mu \nu }$ involves the metric and the affine connection $%
\Gamma _{\ \nu \lambda }^{\mu }$, and sometimes is used as a~third field to
characterize the geometry and codify the gravitational interaction~\cite%
{Adak:2004uh,Adak:2005cd,BeltranJimenez:2017tkd,BeltranJimenez:2019tjy}.
However, both general relativity and teleparallel gravity are based on
 metric connections, which are those that make $Q_{\lambda \mu \nu
}=0$. This metricity requirement implies the validity of the
Leibniz rule for the scalar product of vectors $\mathbf{a}\cdot \mathbf{b}%
=g_{\mu \nu }~a^{\mu }~b^{\nu }$. Therefore, in Equation~\eqref{ortho}, it results
\begin{equation}
(\nabla _{\mathbf{e}_{c}}\mathbf{e}_{a})\cdot \mathbf{e}_{b}+\mathbf{e}%
_{a}\cdot (\nabla _{\mathbf{e}_{c}}\mathbf{e}_{b})=0,
\end{equation}%
where $\nabla _{\mathbf{e}_{c}}=\nabla _{e_{c}^{\lambda }~\partial _{\lambda
}}=e_{c}^{\lambda }\nabla _{\lambda }$. The last equation implies a~simple
property of the spin connection, when~written in an orthonormal basis. In
fact, the affine connection and the spin connection are related in the
following way:\footnote{%
Other authors define the affine connection as $\nabla _{\mathbf{e}_{c}}%
\mathbf{e}_{b}=\Gamma _{cb}^{a}~\mathbf{e}_{a}~$.}

\begin{equation}
\nabla _{\mathbf{e}_{c}}\mathbf{e}_{b}=\Gamma _{bc}^{a}~\mathbf{e}%
_{a},~~~~~\ \ \ ~\ ~\ ~~~~\boldsymbol{\omega }_{\ b}^{a}=\Gamma _{bc}^{a}~%
\mathbf{E}^{c}~~~~\mathrm{or}~~~~~\Gamma _{bc}^{a}=\boldsymbol{\omega }_{\
b}^{a}(\mathbf{e}_{c}).\ 
\end{equation}%

Then, for the orthonormal basis of the previous equation, one gets%
\begin{equation}
0=\Gamma _{ac}^{d}~\mathbf{e}_{d}\cdot \mathbf{e}_{b}+\mathbf{e}_{a}\cdot
\Gamma _{bc}^{d}~\mathbf{e}_{d}=\Gamma _{ac}^{d}~\eta _{db}+\Gamma
_{bc}^{d}~\eta _{ad},
\end{equation}%
and multiplying by $\mathbf{E}^{c}$:%
\begin{equation}
0=\boldsymbol{\omega }_{\ a}^{d}~\eta _{db}+\boldsymbol{\omega }_{\
b}^{d}~\eta _{ad}=\boldsymbol{\omega }_{ba}+\boldsymbol{\omega }_{ab}.
\end{equation}%

Thus, the metricity implies that the spin connection is anti-symmetric in an
orthonormal basis.\footnote{%
The~anti-symmetry of the spin connection implies that $D\eta _{ab}=d\eta
_{ab}-\boldsymbol{\omega }_{\ a}^{c}\eta _{cb}-\boldsymbol{\omega }_{\
b}^{c}\eta _{ac}=-\boldsymbol{\omega }_{ba}-\boldsymbol{\omega }_{ab}=0$,
and $D\varepsilon _{abcd}=0$.} The condition of null torsion and metricity
defines the Levi--Civita connection $\overset{L}{\boldsymbol{\omega }^{a}}_{b} $ in terms of first order derivatives of the tetrad, as a~very well
known formula. It will be convenient for what comes next to define an object
representing the departure of $\boldsymbol{\omega }_{\ b}^{a}$ from the
Levi--Civita connection, which is a~set of 1-forms called the contortion
\begin{equation}
\mathbf{K}_{\ b}^{a}=\boldsymbol{\omega }_{\ b}^{a}-\overset{L}{%
\boldsymbol{\omega }^{a}}_{b}.  \label{K}
\end{equation}%

$\mathbf{K}_{\ b}^{a}$ is a~Lorentz tensor, since any difference of
connections transforms as a~tensor. Notice that $\mathbf{T}^{a}$ and $%
\mathbf{K}_{\ b}^{a}$ relates in a~simple way (use $\overset{L}{\mathbf{T}}%
{}^{a}=0$):%
\begin{equation}
\mathbf{T}^{a}=\mathbf{K}_{\ b}^{a}\wedge \mathbf{E}^{b}.
\end{equation}

The common understanding of the gravitational phenomena assumes that its 
geometrization should be represented by the Levi--Civita spacetime curvature $%
\overset{L}{\mathbf{R}^{a}}_{b}$. Thus, the Einstein--Hilbert Lagrangian is
the 4-form or volume form
\begin{equation}
L_{EH}=\dfrac{1}{4\kappa }~\epsilon _{abcd}\ \mathbf{E}^{a}~\mathbf{E}^{b}%
\overset{L}{\mathbf{R}}{}^{cd}  \label{EHLagr}
\end{equation}%
(we have suppressed the wedge product signs $\wedge $ to abbreviate the
writing, but we must keep in mind that the 1-forms anti-commute). Here, the
tetrad is assumed orthonormal,\ so meaning that a~metric is involved in $%
L_{EH}$. Moreover, $L_{EH}$ is a~Lorentz scalar. Thus, $L_{EH}$ is insensitive
to local boosts and rotations of the tetrad; it is only sensitive to the
metric.

We can use twice the definition of contortion \eqref{K} to relate the
Levi--Civita curvature with the curvature belonging to an arbitrary
connection:
\begin{equation}
\overset{L}{\mathbf{R}}{}^{cd}=\mathbf{R}^{cd}-d\mathbf{K}^{cd}-\mathbf{K}%
_{~e}^{c}~\boldsymbol{\omega }^{ed}-\boldsymbol{\omega }_{~e}^{c}~\mathbf{K}%
^{ed}+\mathbf{K}_{~e}^{c}~\mathbf{K}^{ed}=\mathbf{R}^{cd}-d\mathbf{K}^{cd}-%
\overset{L}{\boldsymbol{\omega }^{c}}_{e}\mathbf{K}^{ed}-\mathbf{K}_{~e}^{c}%
\overset{L}{\boldsymbol{\omega }^{ed}}-\mathbf{K}_{~e}^{c}~\mathbf{K}^{ed},
\end{equation}%
where $-\mathbf{K}_{~e}^{c}\overset{L}{\boldsymbol{\omega }^{ed}}=\overset{L}%
{\boldsymbol{\omega }^{ed}}\mathbf{K}_{~e}^{c}=-\overset{L}{%
\boldsymbol{\omega }^{d}}_{e}\mathbf{K}^{ce}$ (we have used the
anti-commutativity and the metricity). Then,~it~results that%
\begin{equation}
\overset{L}{\mathbf{R}}{}^{cd}=\mathbf{R}^{cd}-\overset{L}{D}\mathbf{K}^{cd}-%
\mathbf{K}_{~e}^{c}~\mathbf{K}^{ed}.
\end{equation}%

Therefore, the Einstein--Hilbert Lagrangian can be written as%
\begin{equation}
L_{EH}=\dfrac{1}{4\kappa }~\epsilon _{abcd}\ \mathbf{E}^{a}~\mathbf{E}%
^{b}\left( \mathbf{R}^{cd}-\overset{L}{D}\mathbf{K}^{cd}-\mathbf{K}_{~e}^{c}~%
\mathbf{K}^{ed}\right).
\end{equation}%

Since the Levi--Civita connection is torsionless, so $\overset{L}{D}\mathbf{E}%
^{a}=0$, we can recognize the boundary term $\overset{L}{D}\left( \epsilon
_{abcd}\ \mathbf{E}^{a}~\mathbf{E}^{b}~\mathbf{K}^{cd}\right) =d\left(
\epsilon _{abcd}\ \mathbf{E}^{a}~\mathbf{E}^{b}~\mathbf{K}^{cd}\right) $ (we
have replaced the covariant exterior derivative with the exterior derivative
since we are differentiating a~Lorentz scalar). Therefore,%
\begin{equation}
L_{EH}=\dfrac{1}{4\kappa }~\epsilon _{abcd}\ \mathbf{E}^{a}~\mathbf{E}%
^{b}\left( \mathbf{R}^{cd}-\mathbf{K}_{~e}^{c}~\mathbf{K}^{ed}\right) -%
\dfrac{1}{4\kappa }d\left( \epsilon _{abcd}\ \mathbf{E}^{a}~\mathbf{E}^{b}~%
\mathbf{K}^{cd}\right).  \label{LEH}
\end{equation}

\subsection{The teleparallel equivalent of general relativity}

Today, it is widely spread that there are at least two other equivalent
ways of codifying gravitational interactions: either by means of the torsion of
spacetime or by its non-metricity. In the teleparallel approach, both the
curvature and the non-metricity are zero; the gravitational phenomena are
encoded in the torsion. Since the spacetime is curvatureless, then the
parallel transport does {not} depend on the path;\ the parallelism is absolute.
There is a~simple way of building a~teleparallel theory equivalent to
general relativity (TEGR): since Equation~\eqref{LEH} is valid whatever the
connection is, we can take a curvatureless connection $\mathbf{R}^{cd}=0$ and
dismiss the boundary term. Thus, we obtain%
\begin{equation}
L_{TEGR}=-\dfrac{1}{4\kappa }~\epsilon _{abcd}\ \mathbf{E}^{a}~\mathbf{E}%
^{b}~\mathbf{K}_{~e}^{c}~\mathbf{K}^{ed}.  \label{LTEGR}
\end{equation}%

This Lagrangian is quadratic in the Levi--Civita connection (i.e., it is
quadratic in first derivatives of the tetrad), and contains a~non-determined
curvatureless spin connection $\boldsymbol{\omega }_{~b}^{a}$. One could
think that the theory itself should govern the connection. This way of
thinking is motivated by general relativity, where the Levi--Civita
connection can be obtained from varying the action independently with
respect to the tetrad and the spin connection (Palatini formalism). In fact,
even if the connection in Equation~\eqref{EHLagr} was not chosen to be $\overset{%
L}{\boldsymbol{\omega }^{a}}_{b}$, we could still obtain it by varying the
Lagrangian with respect to $\boldsymbol{\omega }_{~b}^{a}$:
\begin{equation}
\delta _{\omega }L_{EH}=\dfrac{1}{4\kappa }~\epsilon _{abcd}\ \mathbf{E}^{a}~%
\mathbf{E}^{b}~\delta _{\omega }\mathbf{R}^{cd},
\end{equation}%
where, for any metric connection, it is%
\begin{equation}
\delta _{\omega }\mathbf{R}^{cd}=d\delta \boldsymbol{\omega }^{cd}+\delta %
\boldsymbol{\omega }_{\ e}^{c}~\boldsymbol{\omega }^{ed}+\boldsymbol{\omega }%
_{\ e}^{c}~\delta \boldsymbol{\omega }^{ed}=d\delta \boldsymbol{\omega }%
^{cd}+\boldsymbol{\omega }_{\ e}^{d}~\delta \boldsymbol{\omega }^{ce}~+%
\boldsymbol{\omega }_{\ e}^{c}~\delta \boldsymbol{\omega }^{ed}=D\delta %
\boldsymbol{\omega }^{cd}.
\end{equation}%

Thus, dismissing boundary terms, we obtain%
\begin{equation}
\delta _{\omega }L_{EH}=-\dfrac{1}{4\kappa }~D(\epsilon _{abcd}\ \mathbf{E}%
^{a}~\mathbf{E}^{b})~\delta \boldsymbol{\omega }^{cd}=\dfrac{1}{2\kappa }%
~\epsilon _{abcd}~\mathbf{E}^{a}~\mathbf{T}^{b}~\delta \boldsymbol{\omega }%
^{cd},
\end{equation}%
where we used $D\mathbf{E}^{b}=\mathbf{T}^{b}$. Thus, $\delta _{\omega
}L_{EH}$ is zero if and only if the torsion is zero; therefore, the metric
spin connection must be $\overset{L}{\boldsymbol{\omega }^{a}}_{b}$.

On the other hand, the variation of $L_{TEGR}$ with respect to the
connection,%
\begin{equation}
\delta _{\omega }L_{TEGR}=-\dfrac{1}{4\kappa }~\epsilon _{abcd}\ \mathbf{E}%
^{a}~\mathbf{E}^{b}~\delta _{\omega }(\mathbf{K}_{~e}^{c}~\mathbf{K}^{ed}),
\end{equation}%
where (we will use the metricity)
\begin{equation}
\begin{array}{ll}
\epsilon _{abcd}~\delta _{\omega }(\mathbf{K}_{~e}^{c}~\mathbf{K}^{ed})
&=\epsilon _{abcd}~\delta _{\omega }\mathbf{K}_{~e}^{c}~\mathbf{K}%
^{ed}+\epsilon _{abcd}~\mathbf{K}_{~e}^{c}~\delta _{\omega }\mathbf{K}%
^{ed}=-\epsilon _{abcd}~\mathbf{K}^{ed}~\delta _{\omega }\mathbf{K}%
_{~e}^{c}+\epsilon _{abcd}~\mathbf{K}_{~e}^{c}~\delta _{\omega }\mathbf{K}%
^{ed} \\
&=-\epsilon _{abdc}~\mathbf{K}^{ec}~\delta _{\omega }\mathbf{K}%
_{~e}^{d}+\epsilon _{abcd}~\mathbf{K}_{~e}^{c}~\delta _{\omega }\mathbf{K}%
^{ed}=-\epsilon _{abdc}~\mathbf{K}_{~e}^{c}~\delta _{\omega }\mathbf{K}%
^{ed}+\epsilon _{abcd}~\mathbf{K}_{~e}^{c}~\delta _{\omega }\mathbf{K}^{ed}
 \\
&=2~\epsilon _{abcd}~\mathbf{K}_{~e}^{c}~\delta _{\omega }\mathbf{K}^{ed},
\end{array}
\end{equation}
leads to the result
\begin{equation}
\delta _{\omega }L_{TEGR}=-\dfrac{1}{2\kappa }~\epsilon _{abcd}~\mathbf{E}%
^{a}~\mathbf{E}^{b}~\mathbf{K}_{~e}^{c}~\delta \boldsymbol{\omega }^{ed},
\label{TEGRK}
\end{equation}%
where we used that $\delta _{\omega }\mathbf{K}^{ed}=\delta %
\boldsymbol{\omega }^{ed}$. Noticeably, the vanishing of $\delta _{\omega
}L_{TEGR}$ will not lead to a~curvatureless connection; on the contrary, it
would lead to the Levi--Civita connection again, since the contortion is zero
when $\boldsymbol{\omega }_{\ e}^{c}=\overset{L}{\boldsymbol{\omega }^{c}}%
_{e}$. This shows that the TEGR spin connection cannot be derived from $%
L_{TEGR}$. Instead,~the~curvatureless TEGR spin connection must be chosen,
as it happens with the choice of the torsionless Levi--Civita connection in
the metric formalism of general relativity. The conclusion is that $L_{EH}$
and $L_{TEGR}$ are equivalent from a~\textit{metric} point of view; i.e.,
they are equivalent when regarded as functionals of the tetrad.

Usually, the TEGR spin connection is chosen to be $\boldsymbol{\omega }_{\
e}^{c}=0$ (Weitzenb\"{o}ck connection), which implies that the torsion is $%
\mathbf{T}^{a}=d\mathbf{E}^{a}$. Then, the components of the torsion are
\begin{equation}
\mathbf{T}_{\mu \nu }^{a}=\partial _{\mu }E_{\nu }^{a}-\partial _{\nu
}E_{\mu }^{a},~~~~~~\mathrm{or}~~~~~~T_{~\mu \nu }^{\lambda
}=e_{a}^{\lambda }~(\partial _{\mu }E_{\nu }^{a}-\partial _{\nu }E_{\mu
}^{a}).
\end{equation}%

The affine connection in a~coordinate basis is \footnote{%
Notice that the coefficients $E_{\mu }^{a}~$, $e_{a}^{\mu }$ are the links
between anholonomous and coordinate basis.}%
\begin{equation}
\Gamma _{~\mu \nu }^{\lambda }=e_{a}^{\lambda }~E_{\mu }^{b}~E_{\nu
}^{c}~~\Gamma _{~bc}^{a}+e_{a}^{\lambda }~\partial _{\nu }E_{\mu
}^{a}=e_{a}^{\lambda }~\partial _{\nu }E_{\mu }^{a}.
\end{equation}%

However, in doing so, we have lost one of the fondest properties of the
Lagrangian: if $\boldsymbol{\omega }_{\ e}^{c}$ is fixed to be zero, then $%
L_{TEGR}$ is no longer a~Lorentz scalar. In fact, in Equation~\eqref{LTEGR}, $%
\mathbf{K}_{~e}^{c}$ is a~Lorentz tensor valued 1-form whenever it is a
difference of connections. However, if $\boldsymbol{\omega }_{~e}^{c}=0$,
then $\mathbf{K}_{~e}^{c}$ becomes the connection $\overset{L}{-%
\boldsymbol{\omega }^{c}}_{e}$; it~is no longer a~tensor under local Lorentz
transformations. Therefore, $L_{TEGR}$ stops being a~Lorentz scalar. This
loss of the invariance under local Lorentz transformations of the tetrad is
not a~problem at the level of the dynamical equations. Equation~\eqref{LEH},
where $L_{EH}$ is a~Lorentz scalar, implies that, under local Lorentz
transformations of the tetrad, $L_{TEGR}$ changes by a~boundary term (pseudo-invariance). However,~the~loss of local Lorentz invariance
will be a~real hallmark on modified teleparallel theories like $f(T)$
gravity, which will lead to the appearance of extra degree(s) of freedom \cite{Ferraro:2018tpu}. To
avoid this annoying feature, it has been proposed that $\boldsymbol{\omega }%
_{~e}^{c}$ should not be zero, but belong to the family of the Lorentz
transformations of $\boldsymbol{\omega }_{~b}^{a}=0$,%
\begin{equation}
\boldsymbol{\omega }_{~b}^{a}=\Lambda _{\ a^{\prime }}^{a}~d\Lambda _{\
b}^{a^{\prime }},  \label{inertialSC}
\end{equation}%
where $\Lambda $ represents any Lorentz transformation. This curvatureless form
should not affect the metricity, since~the requirement of metricity is
tensorial in both coordinate indices and Lorentz (tangent space) indices.
For~instance, let us consider infinitesimal Lorentz transformations%
\begin{equation}
\Lambda _{\ a^{\prime }}^{a}=\delta _{\ a^{\prime }}^{a}-\frac{1}{2}~\sigma
^{gh}(x)~(M_{gh})_{\ a^{\prime }}^{a}+O(\sigma ^{2}),  \label{infinit}
\end{equation}%
where $(M_{gh})_{\ a^{\prime }}^{a}$ are the (anti-symmetric) generators of
the vector representation of the Lorentz group,%
\begin{equation}
(M_{gh})_{\ a^{\prime }}^{a}=\delta _{\ g}^{a}~\eta _{ha^{\prime }}-\delta
_{\ h}^{a}~\eta _{ga^{\prime }},  \label{M}
\end{equation}%
and $\sigma ^{gh}(x)$ are anti-symmetric parameters combining boosts and
rotations in the transformation $\Lambda _{\ a^{\prime }}^{a}$.~Then,
\begin{equation}
\begin{array}{ll}
\boldsymbol{\omega }^{ab} &=\eta ^{bc}~\Lambda _{\ a^{\prime
}}^{a}~d\Lambda _{\ c}^{a^{\prime }}=-\frac{1}{2}~\eta ^{bc}~\delta _{\
a^{\prime }}^{a}~(\delta _{\ g}^{a^{\prime }}~\eta _{hc}-\delta _{\
h}^{a^{\prime }}~\eta _{gc})~d\sigma ^{gh}+O(\sigma ^{2})   \\
&=-\frac{1}{2}~(\delta _{\ g}^{a}~\delta _{\ h}^{b}-\delta _{\
h}^{a}~\delta _{\ g}^{b})~d\sigma ^{gh}+O(\sigma ^{2})=-d\sigma
^{ab}+O(\sigma ^{2}).
\end{array}
\end{equation}
Certainly, $\boldsymbol{\omega }^{ab}=-d\sigma ^{ab}$ is metric (although it is
curvatureless only at the first order in $\sigma $).

The claims asserting that the choice \eqref{inertialSC} is the right choice
for what is called ``consistent teleparallel
gravity'' are based on the following argument: since TEGR
is a~theory that is pseudo-invariant under local Lorentz transformations,
there is nothing special about the reference frame on which the spin
connection vanishes, which would mean to put $\Lambda_{b'}^{a}=\delta
_{b'}^{a}$ or any global Lorentz transformation, in \eqref{inertialSC}. This
is entirely true, although completely insubstantial at the level of TEGR. In
fact, the spin connection is not a~dynamical field in TEGR; TEGR is a
dynamical theory just for the metric.

\section{Modified teleparallel gravities}

The traditionally used formulation of TEGR starts from the torsion
scalar $T$ obtained when one replaces $\boldsymbol{\omega }_{~b}^{a}=0$ in
Equation~\eqref{LTEGR}. After some calculation, one gets
\begin{equation}
L_{TEGR}=\frac{1}{2\kappa }~E~T~dx^{0}\wedge dx^{1}\wedge dx^{2}\wedge
dx^{3},
\end{equation}%
where%
\begin{equation}
T\equiv S_{\rho }^{\ \ \mu \nu }\,T_{\ \ \mu \nu }^{\rho },  \label{Tscalar}
\end{equation}%
\begin{equation}
2\ S_{\rho }^{\ \ \mu \nu }\ \equiv \ K_{\ \ \ \rho }^{\mu \nu }+\
T_{\lambda }^{\ \ \lambda \mu }\ \,\delta _{\rho }^{\nu }\ -\ T_{\lambda
}^{\ \ \lambda \nu }\ \,\delta _{\rho }^{\mu },\ 
\end{equation}%
and $K_{\ \ \ \rho }^{\mu \nu }$ results from the components of the
contortion,%
\begin{equation}
K_{\ \ \ \rho }^{\mu \nu }=g^{\nu \lambda }~e_{a}^{\mu }~E_{\lambda
}^{b}~K_{~b\mu }^{a}=\frac{1}{2}\,(T_{\rho }^{\ \ \mu \nu }-T_{\ \ \ \ \rho
}^{\mu \nu }+T_{\ \ \ \ \rho }^{\nu \mu }).
\end{equation}

The torsion scalar in \eqref{Tscalar} is a~suitable starting point for
modifications to general relativity. The~main advantage of this approach is
that nonlinear generalizations of the torsion scalar will always result in
equations of motion of second order in the tetrad field, as $T$ is an object
depending on only first order derivatives. However, linear modifications to
TEGR have been known for a~time. For instance, in~\cite{Hayashi:1979qx}, a gravitational action has been built combining three quadratic pieces
associated with the three irreducible parts of the torsion tensor
(vectorial, axial and traceless-symmetric). The coefficients of the linear
combination are arbitrary, although they are constrained by the physics in
the solar system. However,~they~acquire defined values when local Lorentz
invariance of the theory is required, thus~reducing it to TEGR. On the other
hand, nonlinear modifications of TEGR in the form of arbitrary functions of
the torsion scalar\textit{\ }$T$, better known as $f(T)$ theories of gravity,
have gained attention since they can describe an inflationary early
expansion without resorting to an inflaton field \cite{Ferraro:2006jd}.
In addition, they can mimic dark energy by providing a~late time accelerated
expansion of the universe \cite{Bengochea:2008gz} (for an extensive list of
references, see \cite{Cai:2015emx}).

In $f(T)$ gravity, the action is
\begin{equation}
I=\frac{1}{2\kappa }\int E\,~f(T)~d^{4}x. \label{reltyr}
\end{equation}
Varying this action with respect to the tetrad, the dynamical equations are
obtained:
\begin{equation}
4~e~\partial _{\mu }(f^{\prime }(T)~E~e_{a}^{\lambda }~S_{\lambda }^{\ \mu
\nu })+4~f^{\prime }(T)~e_{a}^{\lambda }~T_{\ \mu \lambda }^{\sigma
}~S_{\sigma }^{\ \mu \nu }-e_{a}^{\nu }~f(T)=-2\kappa ~e_{a}^{\lambda }~%
\mathcal{T}_{\lambda }^{\ \nu },  \label{eqm}
\end{equation}%
where $\mathcal{T}_{\lambda }^{\ \nu }$ is the energy-momentum tensor of
matter coupled to the metric tensor. Originally, $f(T)$ gravity was
conceived for dealing with strong space-time singularities, so high-energy
deformations of the sort $f(T)=T+T^{2}/\lambda +O(1/\lambda ^{2})$ were
considered. For these models, $\lambda $ introduces the length scale $%
\lambda ^{-1/2}$ at which local Lorentz invariance would no longer exist as
a full symmetry, giving way to the remnant ones. For instance, in Ref. \cite%
{Ferraro:2006jd}, the constant $\lambda ^{1/2}$ plays the role of a~maximum
attainable Hubble factor in the context of spatially flat FRW cosmological
models. In that case, the Big Bang singularity is replaced by an early de
Sitter stage of a~purely geometrical character in which the Hubble factor $%
H_{max}=(\lambda /12)^{1/2}$ drives the inflationary era. Much more recently
\cite{Boehmer:2019uxv}, and in a~quite different context, it was shown that
the Schwarzschild curvature singularity is replaced by an infinitely long
cosmic string with constant curvature invariants related to $\lambda$. In
this case, the length scale is $2|\lambda |^{-1/2}$, and it completely
changes the structure of the black hole interior by rendering the spacetime
geodesically complete. In the cosmological example, the~presence of the
scale $H_{max}$ manifests itself at very early times of the cosmic evolution,
while, in the black hole interior, the length scale involves a~constant
curvature asymptotic region located well inside the event horizon. In both
cases, the breaking of the Lorentz symmetry occurs in times or places very
distant from our daily experience.

Since their introduction, $f(T)$ theories have revealed several subtleties
concerning the appearance of preferred (proper) reference frames,
encoded in the tetrad field $\mathbf{E}^{a}(x^{\mu })$. Actually, it was
clear since the early developments in the field that the local action of
the Lorentz group on a~given solution $\mathbf{E}^{a}(x)$ of the $f(T)$
motion equations leads to another tetrad $\mathbf{E}^{a^{\prime }}=\Lambda
_{\ \ a}^{a^{\prime }}(x)\mathbf{E}^{a},$ which is not generally a~solution,
even though both of them generate the same metric tensor $\mathbf{g}=\eta
_{ab}\,\mathbf{E}^{a}\mathbf{E}^{b}$. This is basically because the
equations of motion determine more degrees of freedom than those captured by
the metric tensor~\cite{Ferraro:2018axk}; some attempts have been made for
capturing the number and nature of these degrees of freedom through
Hamiltonian analysis \cite{Ferraro2016,Ferraro:2018tpu}, conformal
transformations \cite{Yang:2010ji,Wright:2016ayu}, cosmological
perturbations \cite{Chen:2010va,Li:2011wu,Izumi:2012qj,Golovnev:2018wbh},
among~others. The~extra degree(s) of freedom define the space-time structure
by means of a~parallelization, which~fixes the tetrad components modulo
certain remnant symmetries associated with the specific solution under
consideration \cite{Ferraro:2014owa}.

\section{Covariance in modified teleparallel gravity}

The covariance of the TEGR action is guaranteed at the level of the
equations of motion, but the Lorentz breaking behavior of the surface term
could be significant for physical quantities defined in the boundary, as
black hole thermodynamics. This issue has been previously investigated \cite%
{Krssak:2015rqa,Krssak:2015lba}; however, in~this section, our focus will be
on discussing covariant versions of modified TEGR. The pseudo-invariant
character of TEGR gravity echoes in nonlinear modifications (as $f(T)$
gravity) in the form of explicit Lorentz breaking at the level of the action
and the equations of motion. This issue has been noticed from the very
beginning \cite{Ferraro:2006jd}, and it was firstly associated with the
appearance of extra degrees of freedom encoded in the components of the
tetrad field. From a~theoretical point of view, covariance under local
Lorentz symmetries is a~desirable feature for any physical theory, but, at
the experimental level, this~symmetry breaking would not be detectable, as
the metric tensor remains unchanged regardless of the privileged orientation
of reference frames. However, the theoretical issue on the covariance of the
Lagrangian formulation is part of an active discussion, and several opinions
have been wielded in the literature. Our aim is to review some of them and
contribute to the debate.

Early discussion on the covariantization of $T$ has been schematized in Ref.
\cite{Sotiriou:2010mv}. The authors consider the possibility of giving up
the teleparallel restriction $\omega _{\ b\nu }^{a}=0$, and take the metric
and the contortion as variables to be independently varied. They conclude
that the $f(T)$ action is dynamically trivial for $K_{\ \mu \nu }^{\rho }$,
since it does not contain derivatives of the contortion, and it is
inconsistent if matter is added. The impossibility of obtaining a
curvatureless connection from varying $T$ with respect to $K_{\ \mu \nu
}^{\rho }$ has been analyzed above (see Equation~\eqref{TEGRK} and below).

An alternative path has been discussed in Ref. \cite{Nester:2017wau}, where
the authors studied a~$f(T)$ action where the covariance of $T$ is restored,
since the \textquotedblleft pure frame\textquotedblright\ approach with $%
\omega _{\ b\nu }^{a}=0$ is not a~priori chosen. Instead, the
curvatureless condition is enforced through Lagrange multipliers in the
action. The advantage of an explicitly covariant Lagrangian is overshadowed
by the introduction of a~large amount of additional fields encoded in these
Lagrange multipliers. Although in principle, in the TEGR case, they describe
the same dynamics because they could be fully determined through the
equations of motion, more research is needed to determine if the nonlinear
generalization can describe equivalent physics as described by ``pure
tetrad'' modified TEGR.

Regarding the covariant formulation in \cite{Krssak:2015oua}, the authors
argue that, if the formulation of $f(T)$ gravity started from a~covariant
version of TEGR, then the equations of motion would acquire an additional
term equal to $f^{\prime }(T)~\omega _{\ a\nu }^{b}S_{b}^{\ \nu \mu }$. This
additional term would allow for recovering the local Lorentz invariance (but
notice that TEGR dynamics are locally invariant without needing this term!).
To obtain this term, the~authors explain that the spin connection %
\eqref{inertialSC} would add a boundary term to the TEGR Lagrangian (this property results from Equation~\eqref{LEH}).  
Since no dynamical equations for the curvatureless spin connection would be obtained in such case, they invoke a~physical criterion to choose it. They say that the spin connection must be \textquotedblleft
inertial\textquotedblright, in the sense that, if gravity is turned off
(i.e. if $\kappa \rightarrow 0$), then the torsion should vanish (so the
torsion would contain just gravitational effects but not inertial ones).
This strategy seems to spoil the purpose of covariantization because the
spin connection is not dynamically determined, but it is found a~posteriori for
each geometry. For a~skeptical mind, this strategy can be considered even
more intricate than sticking to the simplest spin connection choice, that is the
Weitzenb\"{o}ck one. Then, the spin connection becomes an ad hoc
non-dynamical field that removes the undesired effects of an improper
parallelization. Theoretical concerns about the formulation of this covariant approach in a~univocal
standard variational principle come up, in order to avoid introducing
degrees of freedom by~hand.

Some remarks on different possibilities of variational procedures in
teleparallel gravities have been exposed in Ref. \cite{Golovnev:2017dox}.
The case that is of relevance for $f(T)$ gravity is when the variation of
the spin connection is performed in the inertial class. For achieving a
consistent variational procedure, the authors argue that arbitrary Lorentz
matrices $\Lambda _{b}^{a}(x)$ can be considered as the fundamental field
instead of the spin connection. As a~consequence, the TEGR Lagrangian can be
regarded as a~function of both the tetrad and the arbitrary Lorentz matrix,
where the latter now acts as a~physical field too. In the TEGR case, the variation of the
spin connection (that is, the Lorentz matrix) is a~harmless boundary term
that vanishes under proper boundary conditions. In the $f(T)$ case,
however, the variation on the spin connection indeed gives an additional
equation, which reads
\begin{equation}
T_{~\mu \nu }^{\rho }~\partial _{\rho }f^{\prime }(T)+T_{\nu }~\partial
_{\mu }f^{\prime }(T)-T_{\mu }~\partial _{\nu }f^{\prime }(T)=0.
\label{asymcovfT}
\end{equation}%

These are equivalent to the antisymmetric part of the equations of
motion in the pure tetrad approach of $f(T)$ gravity, and its fulfillment is
subject to a~proper parallelization. {This feature also appears in a~wider class of modified teleparallel models} \cite{Hohmann:2017duq}. In \cite{Hohmann:2018rwf} and the
recent review \cite{Krssak:2018ywd}, Equation \eqref{asymcovfT}, written
with an explicit dependence on the inertial spin connection as
\begin{equation}
\partial _{\mu }f^{\prime }(T)~[\partial _{\nu }(E~e_{[a}^{\ \mu }~e_{b]}^{\
\nu })+2~E~e_{c}^{\ [\mu }~e_{[a}^{\ \nu ]}~\omega _{\ b]\nu }^{c}]=0,
\label{asyminertial}
\end{equation}%
has been used in some examples, exhibiting the covariantization procedure.
There are two issues for which we feel the need to provide additional
discussion. The first point has to do with the fact that Equation~\eqref {asyminertial}, and the covariantization procedure, are only possible
whenever the matter contribution in the rhs vanishes. Such contributions,
coming from macroscopic spinorial fluids, have not been properly studied in
the literature due to a~lack of understanding in the correct spinorial
coupling prescription. It~is clear that the covariantization procedure is
not general enough for including this interesting setup, and~further
research is encouraged in this direction, as it could be related with the
physical interpretation of the additional degree(s) of freedom of the $f(T)$
theory.

Secondly, some concerns on the counting of degrees of freedom, and the
first-class constraints that would generate gauge transformations for the
inertial spin connection, have been briefly exposed in \cite{Maluf:2018coz}
(although their criticism is mainly directed to the covariant TEGR
procedure). This resonates with our previous thoughts, and these concerns
are also relevant for the covariantization procedure in $f(T)$ gravity.
Nonetheless, it is relevant to mention that recent work on the Hamiltonian
formalism for the so-called ``new general relativity'' has attempted to
include the inertial spin connection in the formalism~\cite{Blixt:2018znp}.
Their~claim is that the inertial choice in new general relativity represents
pure gauge degrees of freedom, and, consequently, the Weitzenb\"{o}ck
connection is suitable for developing the Hamiltonian analysis (which, of~course, tremendously simplifies the Poisson brackets calculations). It is
clear that the consistency of this approach needs to be revisited, but,
equally important, to analyze if these claims have something to say concerning the
issue of the degree(s) of freedom of $f(T)$ and teleparallel gravity
extensions.

\section{Remnant symmetries and the Lorentz group}

A simple example will help us to fathom some of the subtleties that lie
behind the system \eqref{eqm}. Let us consider the easiest solution of the
vacuum field equations, which should correspond to the absence of gravity.
Thus, let us consider Minkowski spacetime ($Min$), and take the following
tetrad field

\begin{equation}
\mathbf{E}^{0}=dt,\,\,\,\,\mathbf{E}^{1}=dr,\,\,\,\,\mathbf{E}%
^{2}=r\,d\theta ,\,\,\,\,\mathbf{E}^{3}=r\,\sin \theta \,d\phi,
\label{tetradain}
\end{equation}%
which is consistent with the line element $ds^{2}=dt^{2}-dr^{2}-r^{2}(d\theta
^{2}+\sin ^{2}\theta \,d\phi ^{2})$ describing $Min$ in spherical
coordinates $(t,r,\theta ,\phi )$. For the above field, we have $T=2r^{-2}$,
and the equations reduce to the system
\begin{equation}
\begin{array}{rll}
16f_{TT}-2r^{2}f_{T}+fr^{4} &=0,   \\
8f_{TT}-2r^{2}f_{T}+fr^{4} &=0,  \\
-2r^{2}f_{T}+fr^{4} &=0,   \\
f_{TT} &=0,
\end{array}
\end{equation}
where $f_{TT}=f''(T)$, $f_{T} = f'(T)$, and $f=f(T)$. It is clear that the only solution to this system involves $f(T)\propto T$,
which is just GR. Does it mean that $Min$ is not a~solution of vacuum $f(T)$
gravity? What is exactly the problem concerning the frame (\ref{tetradain})?
Let us suppose that, instead of (\ref{tetradain}), we start from
\begin{equation}
\mathbf{E}^{0}=dt,\,\,\,\,\mathbf{E}^{1}=d\rho ,\,\,\,\,\mathbf{E}^{2}=\rho
\,d\phi ,\,\,\,\,\mathbf{E}^{3}=dz,  \label{tetradacil}
\end{equation}%
which leads to the Minkowskian line element $ds^{2}=dt^{2}-d\rho ^{2}-\rho
^{2}d\phi ^{2}-dz^{2}$ written in cylindrical coordinates $(t,\rho ,\phi ,z)$%
. It is simple to check that the Weitzenb\"{o}ck invariant vanishes for the
tetrad \eqref{tetradacil}, and the vacuum field equations \eqref{eqm} take
the remarkably simple form $f(T)=0$. This means that any ultraviolet
deformation of GR (i.e., any function $f(T)$ verifying $f(0)=0$ and $%
f_{T}(0)=1$) admits $Min$ as a~vacuum solution through the fields (\ref%
{tetradacil}).

Finally, consider the (\emph{Euclidean}) tetrad field
\begin{equation}
\mathbf{E}^{0}=dt,\,\,\,\,\mathbf{E}^{1}=dx,\,\,\,\,\mathbf{E}%
^{2}=dy,\,\,\,\,\mathbf{E}^{3}=dz,  \label{tetradacart}
\end{equation}%
which, of course, give us the standard \emph{Cartesian} line element $%
ds^{2}=dt^{2}-\delta _{\alpha \beta }dx^{\alpha }dx^{\alpha }$. Again, $T=0$
and the equations of motion are $f(T)=0$, which are automatically fulfilled
for any UV $f(T)$ deformation of GR.

The point behind the results just obtained (the distinction between
different frames leading to a~certain metric tensor) naturally extends to a
general spacetime $(\mathcal{M},\mathbf{g}(x))$, or, more precisely, $(%
\mathcal{T}^{\star }\mathcal{M},\mathbf{E}^{a}(x))$. We~would like to raise
the following points:
\begin{enumerate}
\item Why, for a~given spacetime $(\mathcal{T}^{\star}\mathcal{M},\mathbf{E}^a(x))$, there are certain \emph{proper} tetrads $\mathbf{E}^a$, and others that definitely do not lead to a~consistent set of equations of motion for a
function $f(T)$ other than the one corresponding to GR?
\item Once the above point was established, is there any way of counting the
number of proper tetrads, and some systematic procedure in order to obtain
them?
\item What is the physical meaning of the proper tetrads?
\end{enumerate}

In relation to the first point, and coming back to the example concerning
Minkowski space, we can find the answer in the topological structure of the
different frames involved. Of course, the manifold topology of $Min$ is just
$R^{4}$, and this is independent from the chosen tetrad. However, at the level
of the vector fields $\mathbf{e}_{a}$, things are different; in the case of
the frame \eqref{tetradain}, we see that every field in the family points at 
one of the directions defined by the spherical coordinates lines, and not
mixed components appear. This means that, as far as the dual vector fields 
concerns, the topology they are actually seeing is $R^{2}\times S^{2}${.
However,}~the~diagonal form of the frame implies that the fields $\mathbf{E}%
^{2}$ and $\mathbf{E}^{3}$ are vector fields covering the whole tangent
space of the 2-sphere, which is impossible because the 2-sphere is not
parallelizable {(spherically symmetric solutions with the same feature have been reported in \cite{Hohmann:2019nat})}. Of~course, the~topological space $R^{2}\times S^{2}$ is
parallelizable, just think about it as $R\times R\times S^{2}$, which is a~product of the parallelizable manifolds $R$ and $R\times S^{2}$.
Nevertheless, the global, everywhere non null, smooth~vectors fields
covering the tangent space of $R^{2}\times S^{2}$, are far from having the
simple diagonal structure of (\ref{tetradain})---see Equation~(\ref{tetminesf})
below.

In the case of the frame (\ref{tetradacil}), the situation is considerably
different. Even though every field in the family is also pointing at 
the
directions defined by the cylindrical coordinates lines, the topology they
are defining is now $R^{3}\times S^{1}$. The diagonal structure suggests that
the parallelization of the entire set is obtained as a~topological product
of parallelizable submanifolds, which is correct in this case due to the
fact that $R$ and $S^{1}$ are parallelizable by themselves. This is the
reason why the fields (\ref{tetradacil}) constitute a~proper tetrad for
describing $Min$.

Finally, the frame (\ref{tetradacart}) also represents a~proper tetrad.
Actually, it should be considered as the canonical, more natural
parallelization of $Min$ because the topology defined by the (Cartesian)
set of vector fields coinciding with one of the latter. Note that both
proper frames (\ref{tetradacil}) and (\ref{tetradacart}) give rise to the
same null value of the Weitzenb\"{o}ck invariant $T$. This is not a
coincidence.

An important observation concerning the role of the local coordinates would
be adequate. Even~though we are adopting local coordinates in order to write
explicit forms of the vector fields, they are not essential at all. Once we
have obtained a~proper tetrad in a~given coordinate system, we can change
coordinates freely. For instance, nothing prevents us from finding a~proper
tetrad for $Min$ in spherical coordinates; just take (\ref{tetradacil}) or (%
\ref{tetradacart}), and change coordinates accordingly; we will find $E_{\mu
^{\prime }}^{a}=\partial x^{\mu }/\partial x^{\mu ^{\prime }}\,E_{\mu }^{a}$%
, where $x^{\mu }$ refers to the cylindrical or Cartesian chart (depending
on the case), and $x^{\mu ^{\prime }}$ to the spherical one. If we start
from the canonical frame (\ref{tetradacart}), we will get
\begin{equation}
\begin{array}{lll}
\mathbf{E}^{0} &=dt,   \\
\mathbf{E}^{1} &=\sin\theta \cos\phi \ dr+r\cos \theta \cos \phi \ d\theta
-r\sin \theta \sin \phi \ d\phi ,  \\
\mathbf{E}^{2} &=\sin \theta \sin \phi \ dr+r\cos \theta \sin \phi \
d\theta +r\sin \theta \cos \phi \ d\phi,  \\
\mathbf{E}^{3} &=\cos\theta \ dr-r\sin \theta \ d\theta.  \label{tetminesf}

\end{array}
\end{equation}%

Clearly, a~parallelization of $Min$ in spherical coordinates is quite more
involved than the (incorrect) choice~(\ref{tetradain}). Take note that the
block structure in (\ref{tetminesf}) emphasizes the fact that the spatial part of
the frame is actually a~parallelization of the submanifold $R \times S^{2}$.


\subsection{The remnant group}

\label{sec2} We devote this section to the second point raised above. A
systematic approach to the study of the proper frames describing a~certain
spacetime $(\mathcal{T}^{\star}\mathcal{M},\mathbf{E}^a(x))$ within the
context of $f(T)$ gravity involves the characterization of the \emph{%
remnant group} of Lorentz transformations of $(\mathcal{T}^{\star}\mathcal{M}%
,\mathbf{E}^a(x))$. Details of the following exposition can be found in \cite%
{Ferraro:2014owa}.

In general, under a~Lorentz transformation of the tetrad $\mathbf{E}%
^{a}\rightarrow \mathbf{E}^{a^{\prime }}=\Lambda _{\ \ a}^{a^{\prime }}%
\mathbf{E}^{a}$, the torsion scalar $T$ transform is as follows:
\begin{equation}
T~\mathbf{\Omega }\rightarrow T^{^{\prime }}\mathbf{\Omega }=T~\mathbf{%
\Omega ~}+~d(\epsilon _{abcd}\ \mathbf{E}^{a}\wedge \mathbf{E}^{b}\wedge
\eta ^{de}\Lambda _{\ \ f^{\prime }}^{c}\ d\Lambda _{\ e}^{f^{\prime }})
\label{invariance1}
\end{equation}%
($\mathbf{\Omega }=E~dx^{0}\wedge dx^{1}\wedge dx^{2}\wedge dx^{3}~$is the
spacetime volume), so the torsion scalar is not really a~Lorentz scalar. The
remnant group $\mathcal{A}(\mathbf{E}^{a})$ of a~given spacetime $(\mathcal{T%
}^{\star }\mathcal{M},\mathbf{E}^{a}(x))$ is defined as the subgroup of $%
SO(1,3)$ under which $T$ becomes a~Lorentz scalar, i.e., by demanding
\begin{equation}
d(\epsilon _{abcd}\ \mathbf{E}^{a}\wedge \mathbf{E}^{b}\wedge \eta
^{de}\Lambda _{\ \ f^{\prime }}^{c}\ d\Lambda _{\ e}^{f^{\prime }})=0.
\label{invariance}
\end{equation}%

Of course, the global Lorentz group ($d\Lambda _{\ e}^{f^{\prime }}=0$) is
always included in $\mathcal{A}(\mathbf{E}^{a})$, for all $\mathbf{E}^{a}$.
For infinitesimal transformations, we have the expression (\ref{infinit}),
where the matrices $M_{gh}$ given in Equation~(\ref{M}) are boosts generators, $%
K_{\alpha }=M_{0\alpha }$, and rotation generators $J_{\alpha }=-\frac{1}{2}%
\epsilon _{\alpha \beta \gamma }\,M^{\beta \gamma }$. The Lorentz algebra
reads%
\begin{equation}
\lbrack J_{\alpha },J_{\beta }]=\epsilon _{\alpha \beta \gamma }\ J^{\gamma
},\,\,\,\,\,\,\,\,[K_{\alpha },K_{\beta }]=-\epsilon _{\alpha \beta \gamma
}\ J^{\gamma },\,\,\,\,\,\,\,\,[K_{\alpha },J_{\beta }]=\epsilon _{\alpha
\beta \gamma }\ K^{\gamma }.  \label{algebrakj}
\end{equation}%

It is straightforward to show that Equation~(\ref{invariance}) at the
infinitesimal level reduces to
\begin{equation}
\epsilon _{abcd}\ d(\mathbf{E}^{a}\wedge \mathbf{E}^{b})\wedge d\sigma
^{cd}\,=\,0.\   \label{condition}
\end{equation}%

At this point, it results in being very convenient to introduce a~useful concept with
regard to the solutions of Equation~(\ref{condition}). A sufficient
condition for a~given element of $SO(3,1)$ to belong to $\mathcal{A}(E^{a})$%
comes when one classifies the solutions $\mathbf{E}^{a}$ of the equations
of motion (\ref{eqm}) according to the number of closed 2-forms they
involve. A solution $\{\mathbf{E}^{a}\}$ will be called an \emph{%
n-closed-area frame} ($n$-CAF), if it satisfies \mbox{$d(\mathbf{E}^{a}\wedge
\mathbf{E}^{b})=0$} for $n$ of the six different pairs $(a,b)$. The utility of
this definition can be seen through a~couple of simple examples.

Let us suppose that a~given solution has the 1-CAF defined by the 2-form $%
\mathbf{E}^{0}\wedge \mathbf{E}^{1}$. As $d(\mathbf{E}^{0}\wedge \mathbf{E}%
^{1})=0$, the local parameter $\sigma ^{23}$ is totally free in Equation (\ref%
{condition}), which means that $M_{23}\equiv J_{1}$ can be freely chosen. In
other words, we have that $Rot_{x^{1}}\subset \mathcal{A}(E^{a})$ (here $%
Rot_{x^{1}}$ stands for the subgroup of rotations about the $x^{1}$-axis).
Albeit $\{\mathbf{E}^{a}\}$ is only a~1-CAF, other remnant symmetries could
be allowed. Imagine that

\begin{equation}
d(\mathbf{E}^{1}\wedge \mathbf{E}^{2})\propto dt\wedge dx^{1}\wedge dx^{2}.
\label{simrem}
\end{equation}%

This means that a~Lorentz boosts $\sigma ^{03}(t,~x^{1},~x^{2})$ will also
be in $\mathcal{A}(E^{a})$ (note the restriction on the coordinates in the
boost generator). This is so because {the} 1-form $d\sigma ^{03}$ in Equation~(\ref%
{condition}) does not contain a~term proportional to $dx^{3}$, so the wedge
product $d(\mathbf{E}^{1}\wedge \mathbf{E}^{2})\wedge d\sigma ^{03}=0$. In
this way, not only are the rotations about the $x^{1}-$axis in $\mathcal{A}%
(E^{a})$, but also local (restricted) Lorentz boosts along the $x^{3}-$%
direction are admissible.

The structure of $\mathcal{A}(E^{a})$ becomes richer as the number of closed
areas approach its maximum value $n=6$. In general, for an $n$-CAF, we
should expect $n$ one-parameter subgroups, but two-parameter Abelian
subgroups could also exist for $n\geq2$; take, for instance, the 2-CAF
defined by $d(\mathbf{E}^{0}\wedge\mathbf{E}^{3})=d(\mathbf{E}^{1}\wedge%
\mathbf{E}^{2})=0$. This means that the remnant group will include local
transformations generated by combinations of $M_{12}$ and $M_{03}$, which
lead us two the 2-parameter Abelian group $\{J_{3},K_{3}\}$. More details on
the structure of $\mathcal{A}(E^{a})$ can be consulted in \cite%
{Ferraro:2014owa}.

Before proceeding to show specific examples, we should
mention the following crucial fact; according~to Equation~(\ref{condition}), if $%
\{\mathbf{E}^{a}\}$ is a~6-CAF, then we get the maximum remnant symmetry
because all the infinitesimal parameters $\sigma ^{ab}$ remain free. Then, we
obtain
\begin{equation}
SO(3,1)_{inf}\subset \mathcal{A}(E^{a}),
\end{equation}%
where $SO(3,1)_{inf}$ represents the infinitesimal Lorentz group. This
result will be of {utmost} importance concerning the physical interpretation
of $\mathcal{A}(E^{a})$.

Moving on now to a~more global level, let us study the importance of $%
\mathcal{A}(E^{a})$ in solutions describing cosmological models. Let us
consider anisotropic, homogeneous Bianchi type I spacetimes given by the
line element
\begin{equation}
ds^{2}=dt^{2}-a_{1}^{2}(t)\,dx^{2}-a_{2}^{2}(t)\,dy^{2}-a_{3}^{2}(t)\,dz^{2},
\label{metcurv}
\end{equation}%
where we used Euclidian coordinates $(t,x,y,z)$ and $a_{i}(t)$ are the scale
factors. Metric (\ref{metcurv}) contains the important case $%
a_{1}=a_{2}=a_{3}$ representing a~spatially flat FRW model. The topology of $%
(\mathcal{M},\mathbf{g}(x))$ is $R^{4}$, then a~canonical parallelization
can be easily obtained by means of
\begin{equation}
\mathbf{E}^{0}=dt,\,\,\,\mathbf{E}^{1}=a_{1}(t)\,dx,\,\,\,\mathbf{E}%
^{2}=a_{2}(t)\,dy,\,\,\,\mathbf{E}^{3}=a_{3}(t)\,dz.  \label{frame3TAC}
\end{equation}%

Some clue on the structure of $\mathcal{A}(E^{a})$ can be obtained by
realizing that $d(\mathbf{E}^{0}\wedge \mathbf{E}^{\alpha })=0$, $\forall
\alpha $, so the tetrad (\ref{frame3TAC}) is a~3-CAF. This automatically
implies that any local rotation $\sigma ^{\beta \gamma }(x)$ of the frame (%
\ref{frame3TAC}) will be an admissible symmetry. This is so because, for
every pair $(0,\alpha )$, we have a~pair $(\beta ,\gamma )$ ($\alpha \neq
\beta \neq \gamma $), and there are three such pairs. Thus, despite the fact
that metric (\ref{metcurv}) depends solely on time, proper tetrads depending
on all the coordinates can be obtained by locally rotating the canonical
frame (\ref{frame3TAC}).

Curiously enough, there are many more remnant symmetries present in this
spacetime. Even though $\mathbf{E}^{\alpha }\wedge \mathbf{E}^{\beta }$
is not closed, we have
\begin{equation}
d(\mathbf{E}^{\alpha }\wedge \mathbf{E}^{\beta })=(a_{\alpha }a_{\beta
})_{,\,t}\,\,dt\wedge dx^{\alpha }\wedge dx^{\beta },  \label{cosmo3}
\end{equation}
{where $\alpha, \beta$ are fixed indices.} This is a~nice physical example of additional (restricted) symmetries (see
the discussion following Equation~(\ref{simrem}) in the previous section). As
explained opportunely, the restriction on the spatial coordinates comes from
the boost generator $\sigma ^{0\gamma }(t,x^{\alpha },x^{\beta })$, $\alpha
\neq \beta \neq \gamma $. Because of this, the~structure of $\mathcal{A}%
(E^{a})$ is very rich and includes the two-parametric Abelian groups
\begin{equation}
\begin{array}{lll}  \label{restricted}
&\{K_{x}(t,y,z),J_{x}(x^{\mu})\},  \\
&\{K_{y}(t,x,z),J_{y}(x^{\mu})\},   \\
&\{K_{z}(t,x,y),J_{z}(x^{\mu})\},
\end{array}
\end{equation}
where it is not in vain to emphasize again the coordinate restriction on the
boost generators. The~conclusions just obtained are also valid for the
spatially flat FRW models, which seem to be in good agreement with the
experimental facts concerning the large scale structure of the Universe.

Considering the nature of the space in question, we see that the remnant
symmetries still constitute a~vast, overrated set. In any physical
cosmological model, it is always assumed that non-privileged positions exist;
this assumption is of course supported by the fact that the Universe seems
to look the same by looking at any direction, and nothing seems to indicate
that this is valid only for special points in it, us between them. Of
course, local irregularities such as stars or galaxies exist here and there, but
these are considered tiny particles that are distributed in an
approximately isotropic and homogeneous way throughout space. Thus, why bother with transformations depending on spatial coordinates? Why even consider them if here and there are basically the same?
In other words, why demand local symmetries that are not consistent with
the type of physical situations we are trying to model?

In this sense, the remnant symmetries are representative of the solution
under consideration, and not general properties of all spaces. A
cosmological spatially flat FRW model, for instance, is a~physical theory
about the temporal evolution of the scale factor $a(t)$, or, more
properly, about the Hubble rate $H(t)$. A pair of observers connected by
Lorentz transformations should be able to measure, let us say, the value of $%
H$. This will enable them to infer, by means of the equations of motions,
physical properties of the Universe, as the distribution of energy-matter.
However, from the outset, the theory predicts (for a~given moment of time) the
same value of $H(t)$ irrespective of the spatial position of the observers.
Thus,~only~time dependent boost or rotations connecting physical observers
should be considered as strictly physical symmetries in this specific model.
As shown in Equation~(\ref{restricted}), $\mathcal{A}(E^{a})$ includes these
transformations. Any~other dependence of the local parameters is simply
asking too much.

\section{Conclusions}

We have summarized several arguments about covariance in modified
teleparallel gravities. There~are still several issues to be addressed, in
particular regarding the coupling of the antisymmetric part of the
equations of motion of modified teleparallel to spinorial sources. It is of
utmost importance to obtain a~variational principle for this approach.
Considerations about the inheritance of a~remnant symmetry group of local
Lorentz transformations have been discussed, which present an intriguing
structure.

\vspace{6pt}

\section*{Acknowledgments} The authors thank Christian G. Boehmer, Alan A. Coley, Alexey Golovnev, Manuel Hohmann, Robert J. van den Hoogen and Martin Kr\v{s}\v{s}\'{a}k for comments and helpful discussion. C.B. and M.J.G. are grateful to the organizers of the conference ``Teleparallel Universes in Salamanca'' for the kind invitation. This work was partially supported by the Consejo Nacional de Investigaciones Cient\'ificas y T\'ecnicas (CONICET), Universidad de Buenos Aires and Instituto Balseiro
(UNCUYO). M.J.G. has been funded by CONICYT-FONDECYT Postdoctoral grant N$^o$3190531. C.B., R.F., and F.F. are members of Carrera del Investigador Cient\'ifico. 


\end{document}